\documentclass[physrev,reprint,nofootinbib,superscriptaddress]{revtex4-1}

\usepackage{graphicx}
\usepackage{dcolumn}
\usepackage{bm}
\usepackage{amsmath}
\usepackage{amssymb}
\usepackage{booktabs}
\usepackage{multirow}
\usepackage{physics}
\usepackage{xcolor}
\usepackage{siunitx}
\usepackage{afterpage}
\usepackage{mleftright}

\usepackage{xtab}
\usepackage[multiple]{footmisc}
\usepackage{tcolorbox}

\usepackage{pdfpages}
\usepackage{etoolbox}

\usepackage[left=20mm,right=20mm,top=25mm,bottom=25mm,a4paper]{geometry}

\makeatletter
\patchcmd{\@outputpage@head}{\@ifx{\LS@rot\@undefined}{}{\LS@rot}}{}{}{}
\newcommand*{\balancecolsandclearpage}{%
  \close@column@grid
  \cleardoublepage
  \twocolumngrid
\newcommand\blankpage{%
  \null
  \thispagestyle{empty}%
  \addtocounter{page}{-1}%
  \newpage}
}
\makeatother

\protected\def\verythinspace{
  \ifmmode
    \mskip0.5\thinmuskip
  \else
    \ifhmode
      \kern0.08334em
    \fi
  \fi
}

\makeatletter
\renewcommand\@make@capt@title[2]{%
 \@ifx@empty\float@link{\@firstofone}{\expandafter\href\expandafter{\float@link}}%
  {\textbf{#1: }}#2\quad
}%
\makeatother

\usepackage{url}

\usepackage[colorlinks]{hyperref}
\hypersetup{%
        plainpages=true,
        breaklinks=true,
        hypertexnames=false,
        pageanchor=true,
        colorlinks=true,
        linkcolor={blue},
        citecolor={blue},
        urlcolor={blue},
        anchorcolor={black}
      }

\begin{document}

\title{Entangling Schr{\"o}dinger's cat states by bridging discrete- and continuous-variable encoding}

\author{Daisuke Hoshi}
\thanks{These authors contributed equally to this work.\\email: kwon2866@gmail.com (Sangil Kwon)}
\affiliation{Department of Physics, Graduate School of Science, Tokyo University of Science, 1-3 Kagurazaka, Shinjuku-ku, Tokyo 162-8601, Japan}
\affiliation{RIKEN Center for Quantum Computing (RQC), Wako-shi, Saitama 351-0198, Japan}

\author{Toshiaki Nagase}
\thanks{These authors contributed equally to this work.\\email: kwon2866@gmail.com (Sangil Kwon)}
\affiliation{Department of Physics, Graduate School of Science, Tokyo University of Science, 1-3 Kagurazaka, Shinjuku-ku, Tokyo 162-8601, Japan}
\affiliation{RIKEN Center for Quantum Computing (RQC), Wako-shi, Saitama 351-0198, Japan}

\author{Sangil Kwon}
\thanks{These authors contributed equally to this work.\\email: kwon2866@gmail.com (Sangil Kwon)}
\affiliation{Research Institute for Science and Technology, Tokyo University of Science, 1-3 Kagurazaka, Shinjuku-ku, Tokyo 162-8601, Japan}

\author{Daisuke Iyama}
\affiliation{Department of Physics, Graduate School of Science, Tokyo University of Science, 1-3 Kagurazaka, Shinjuku-ku, Tokyo 162-8601, Japan}
\affiliation{RIKEN Center for Quantum Computing (RQC), Wako-shi, Saitama 351-0198, Japan}

\author{Takahiko Kamiya}
\affiliation{Department of Physics, Graduate School of Science, Tokyo University of Science, 1-3 Kagurazaka, Shinjuku-ku, Tokyo 162-8601, Japan}
\affiliation{RIKEN Center for Quantum Computing (RQC), Wako-shi, Saitama 351-0198, Japan}

\author{Shiori Fujii}
\affiliation{Department of Physics, Graduate School of Science, Tokyo University of Science, 1-3 Kagurazaka, Shinjuku-ku, Tokyo 162-8601, Japan}
\affiliation{RIKEN Center for Quantum Computing (RQC), Wako-shi, Saitama 351-0198, Japan}

\author{Hiroto Mukai}
\affiliation{RIKEN Center for Quantum Computing (RQC), Wako-shi, Saitama 351-0198, Japan}
\affiliation{Research Institute for Science and Technology, Tokyo University of Science, 1-3 Kagurazaka, Shinjuku-ku, Tokyo 162-8601, Japan}

\author{Shahnawaz Ahmed}
\affiliation{Department of Microtechnology and Nanoscience, Chalmers University of Technology, 412 96 Gothenburg, Sweden}

\author{Anton Frisk Kockum}
\affiliation{Department of Microtechnology and Nanoscience, Chalmers University of Technology, 412 96 Gothenburg, Sweden}

\author{Shohei Watabe}
\affiliation{Research Institute for Science and Technology, Tokyo University of Science, 1-3 Kagurazaka, Shinjuku-ku, Tokyo 162-8601, Japan}
\affiliation{College of Engineering, Shibaura Institute of Technology, 3-7-5 Toyosu, Koto-ku, Tokyo 135-8548, Japan}

\author{Fumiki Yoshihara}
\affiliation{Department of Physics, Graduate School of Science, Tokyo University of Science, 1-3 Kagurazaka, Shinjuku-ku, Tokyo 162-8601, Japan}
\affiliation{Research Institute for Science and Technology, Tokyo University of Science, 1-3 Kagurazaka, Shinjuku-ku, Tokyo 162-8601, Japan}

\author{Jaw-Shen Tsai}
\affiliation{RIKEN Center for Quantum Computing (RQC), Wako-shi, Saitama 351-0198, Japan}
\affiliation{Research Institute for Science and Technology, Tokyo University of Science, 1-3 Kagurazaka, Shinjuku-ku, Tokyo 162-8601, Japan}
\affiliation{Graduate School of Science, Tokyo University of Science, 1-3 Kagurazaka, Shinjuku-ku, Tokyo 162-8601, Japan}


\date{\today}

\begin{abstract}
In quantum information processing, two primary research directions have emerged:
one based on discrete variables (DV) and the other on the structure of quantum states in a continuous-variable (CV) space.
Integrating these two approaches could unlock new potentials, overcoming their respective limitations.
Here, we show that such a DV--CV hybrid approach, applied to superconducting Kerr parametric oscillators (KPOs), enables us to entangle a pair of Schr{\"o}dinger's cat states by two methods.
The first involves the entanglement-preserving conversion between Bell states in the Fock-state basis (DV encoding) and those in the cat-state basis (CV encoding).
The second method implements a $\sqrt{\textrm{iSWAP}}$ gate between two cat states following the procedure for Fock-state encoding.
This simple and fast gate operation completes a universal quantum gate set in a KPO system. 
Our work offers powerful applications of DV--CV hybridization and marks a first step toward developing a multi-qubit platform based on planar KPO systems.
\end{abstract}

\maketitle

\section{Introduction}

\begin{figure*}
\centering
\includegraphics[scale=0.95]{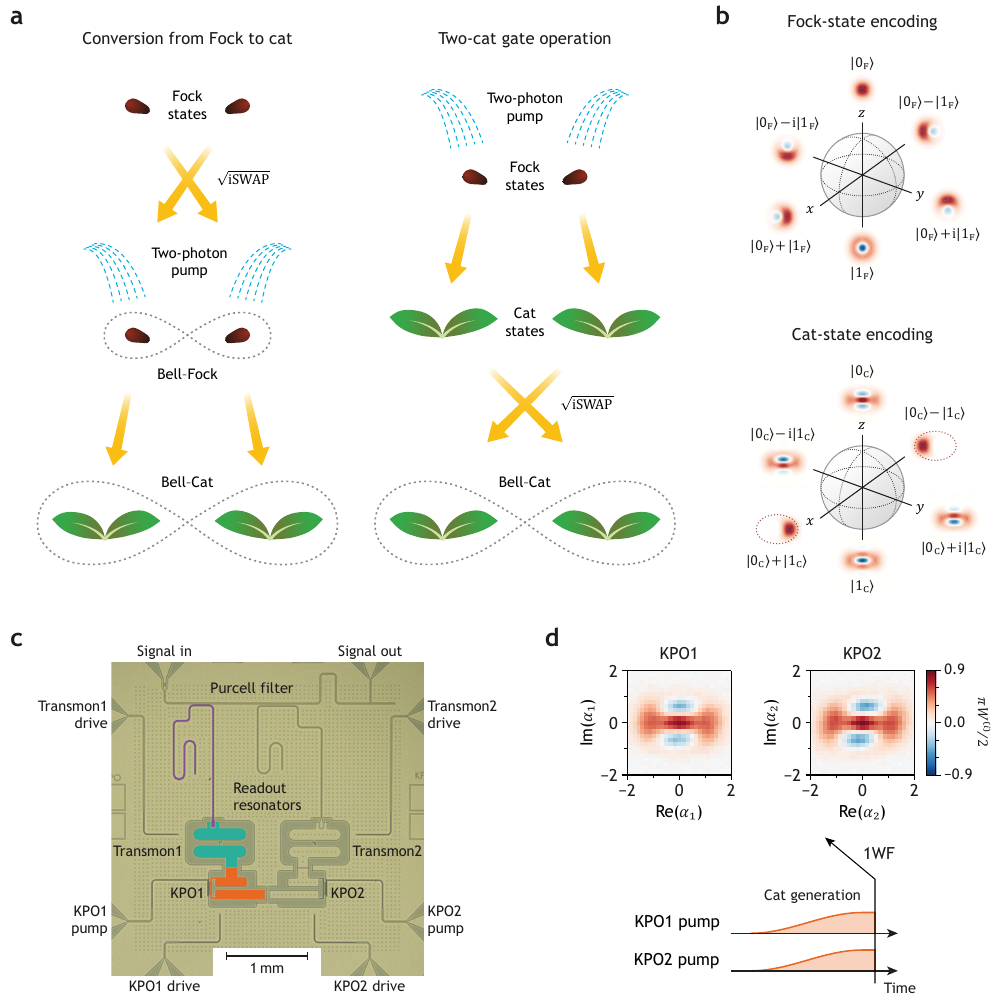}
\caption{\textbf{Concept of the experiment.}
\textbf{a} Seed-sprout analogy for our methods to create Bell--Cat states.
In this analogy,
the seeds represent Fock state encoding,
the sprouts represent cat state encoding,
water sprinkles represent two-photon pumps, and
gray dotted curves indicate entanglement.
\textbf{b} Bloch spheres for Fock state encoding and cat state encoding.
The normalization factor was omitted for simplicity.
\textbf{c} Figure of the chip. The left side is in false colour for clarity.
Each KPO is composed of 10 direct-current superconducting quantum interference devices (DC SQUIDs) with a shunting capacitor.
The two KPOs are capacitively coupled.
The state of each KPO is monitored by the nearby transmon (green) and its readout resonator (purple).
\textbf{d} Simultaneous and independent generation of even cat states $\ket{0_\textrm{C}}$ from vacuum states $\ket{0_\textrm{F}}$ and the corresponding pulse sequence.
The pulse sequence used to measure the Wigner function is omitted for simplicity (see Supplementary Fig.~2 for the full pulse sequence).
The colour represents the scaled one-mode Wigner function (1WF), i.e., the number parity.
}
\label{fig:sprout}
\end{figure*}

For nearly three decades, there have been two paradigms in quantum information processing:
one involves discrete variables (DVs), such as photon number (Fock) states or spin states \cite{vandersypen, mit, kwon, burkard},
whereas the other relies on the structure of quantum states in a continuous-variable (CV) space, such as Schr{\"o}dinger's cat and Gottesman--Kitaev--Preskill states \cite{braunstein2005, joshi2021, eriksson2024}.
Recently, considerable efforts have focused on bridging DV and CV quantum information to overcome the limitations of each paradigm \cite{andersen2015, jeong2014, morin2014, ulanov2017, sychev2018, gan2020, darras2023, macridin2024}.
Parametrically driven Kerr nonlinear resonators, often referred to as Kerr parametric oscillators (KPOs) \cite{dykman, goto2019b, wustmann2019, yamaji2022, yamaguchi2024}, offer a unique testbed for this task, particularly for exploring emergent quantum properties like entanglement in interacting quantum systems.
This capability is enabled by simple one-to-one conversion between Fock and cat states via parametric pump control \cite{cochrane1999, goto2016a, minganti2016, puri2017a, zhang2017, wang2019, masuda2021a, xue2022}.

In our previous work \cite{catGen}, we experimentally demonstrated that such conversion in a superconducting planar KPO preserves the quantum coherence of the system, with the underlying physics being quantum tunnelling in phase space \cite{marthaler2007, venkatraman2022}.
Furthermore, we showed that single-gate operations on cat states in a KPO can be implemented similarly to conventional gate operations on the Fock-state basis \cite{goto2016b, puri2020, grimm2020, kanao2021b, xu2021, masuda2022, hajr2024}.

To establish KPO systems as a promising quantum information platform, the next step would be extending our approach to a multi-KPO system.
Although there have been studies on two interacting KPOs \cite{yamaji2023, margiani2023, alvarez2024}, the entanglement between them and its preservation during the conversion between Fock and cat states have yet to be investigated.
Additionally, the two-KPO gate operation for cat-state encoding, which we refer to as the two-cat gate, has not been demonstrated.

In this work, we introduce two straightforward methods to create entangled cat states---a valuable resource for fault-tolerant quantum computation and communication \cite{DSI, sanders2012, walschaers2021, wang2016, albert2019, zhou2021, gertler2023}---by bridging DV and CV domains.
The first method is the entanglement-preserving and deterministic conversion from Fock-state encoding to cat-state encoding.
Such a conversion suggests the possibility of constructing quantum networks in the cat basis using conventional schemes originally developed for the Fock basis, thereby reducing experimental complexity.
Thus, our demonstration highlights the potential of DV--CV hybridization and may lay new groundwork for constructing quantum networks in the cat basis.

The next method is to implement a $\sqrt{\textrm{iSWAP}}$ gate between two cat states in a manner almost identical to that for Fock-state encoding \cite{chono2022}.
This allows us to create entangled cat states faster than previous implementations on bosonic modes \cite{gao2019, chapman2023}, using only a single square pulse.
Our implementation completes the demonstration of a universal quantum gate set, alongside the single-cat gate operations from our previous work \cite{catGen}.

For both our methods, we can make analogies to seeds (from the DV domain) sprouting (in the CV domain) thanks to watering (two-photon pumping), as illustrated in Fig.~\ref{fig:sprout}a.
In this paper, we denote Fock states $\ket{0}$ and $\ket{1}$ as $\ket{0_\textrm{F}}$ and $\ket{1_\textrm{F}}$, respectively.
Correspondingly, the even and odd cat states are denoted as $\ket{0_\textrm{C}}$ and $\ket{1_\textrm{C}}$ as shown in Fig.~\ref{fig:sprout}b.
In addition, we refer to the Bell states in the Fock basis as Bell--Fock states and designate the resulting entangled cat states as Bell--Cat states.

\section{Results}

\subsection{Setup}

The chip used in this work is shown in Fig.~\ref{fig:sprout}c.
It is the same chip used in our previous study \cite{catGen}.
The transition frequencies between the $\ket{0}$ and $\ket{1}$ states of the KPOs are 2.564 GHz (KPO1) and 2.420 GHz (KPO2).
The self-Kerr coefficient of both KPOs is approximately 2 MHz after ramping up the pump.

The Hamiltonian of our system can be described as
(see Sec.~1 of Supplementary Information for the derivation)
\begin{align}
\begin{split}\label{eq:twoKPO}
\hat{\mathcal{H}}(t) =\,\verythinspace&
\Delta_1 \hat{a}_1^\dagger\hat{a}_1 
- \frac{K_1}{2} \hat{a}_1^\dagger\hat{a}_1^\dagger \hat{a}_1\hat{a}_1
+ \frac{P_1(t)}{2} \mleft( \hat{a}_1^\dagger\hat{a}_1^\dagger + \hat{a}_1\hat{a}_1 \mright) \\
+\,& \Delta_2 \hat{a}_2^\dagger\hat{a}_2 
- \frac{K_2}{2} \hat{a}_2^\dagger\hat{a}_2^\dagger \hat{a}_2\hat{a}_2
+ \frac{P_2(t)}{2} \mleft( \hat{a}_2^\dagger\hat{a}_2^\dagger + \hat{a}_2\hat{a}_2 \mright) \\
+\,& g \mleft(
\hat{a}_1^\dagger\hat{a}_2 \textrm{e}^{+\textrm{i}\Delta_\textrm{p}t}
+ \hat{a}_1\hat{a}_2^\dagger \textrm{e}^{-\textrm{i}\Delta_\textrm{p}t}
\mright).
\end{split}
\end{align}
Here, we are working in units where $\hbar=1$;
$\hat{a}_i$ and $\hat{a}_i^\dagger$ are the ladder operators for the KPO$i$ $(i=1,2)$;
$\Delta_i (\equiv \omega_{\textrm{K}i} - \omega_{\textrm{p}i}/2)$
is the KPO-pump frequency detuning, where
$\omega_{\textrm{K}i}$ is the transition frequency between the $\ket{0_\textrm{F}}$ and $\ket{1_\textrm{F}}$ states, and $\omega_{\textrm{p}i}$ is the frequency of the two-photon pump;
$K_i$ is the self-Kerr coefficient;
$P_i$ is the amplitude of the pump;
$g$ is the coupling constant; and
$\Delta_\textrm{p} [\equiv\!(\omega_\textrm{p1}-\omega_\textrm{p2})/2]$ is half of the detuning between the two pumps.
The Hamiltonian in Eq.~\eqref{eq:twoKPO} is in the rotating frame defined by $\hat{\mathcal{H}}_0 = \sum_i (\omega_{\textrm{p}i}/2)\hat{a}_i^\dagger\hat{a}_i$.
See Supplementary Table~1 for the values of these system parameters.

The cat states are generated adiabatically using the pump pulse with the profile $\sin^2(\pi t/2\tau_\textrm{ramp})$, where the ramping time $\tau_\textrm{ramp}$ is 1 \unit{\micro\second} (see Methods for more details).
Throughout this work, for both KPOs, the $P/K$ ratio is chosen to be 1.0, and the pump detuning [$\Delta_1$ and $\Delta_2$ in Eq.~\eqref{eq:twoKPO}] is chosen to be 1.0 MHz.
Since the detuning between the two KPOs (144 MHz) is nearly 20 times larger than the coupling (8 MHz), the interaction is effectively turned off on the timescale of the measurements;
thus, cat states can be generated and measured independently and simultaneously as shown in Fig.~\ref{fig:sprout}d.

\subsection{Conversion from Fock to cat}
\label{sec:conversion}

\begin{figure*}
\centering
\includegraphics[scale=0.95]{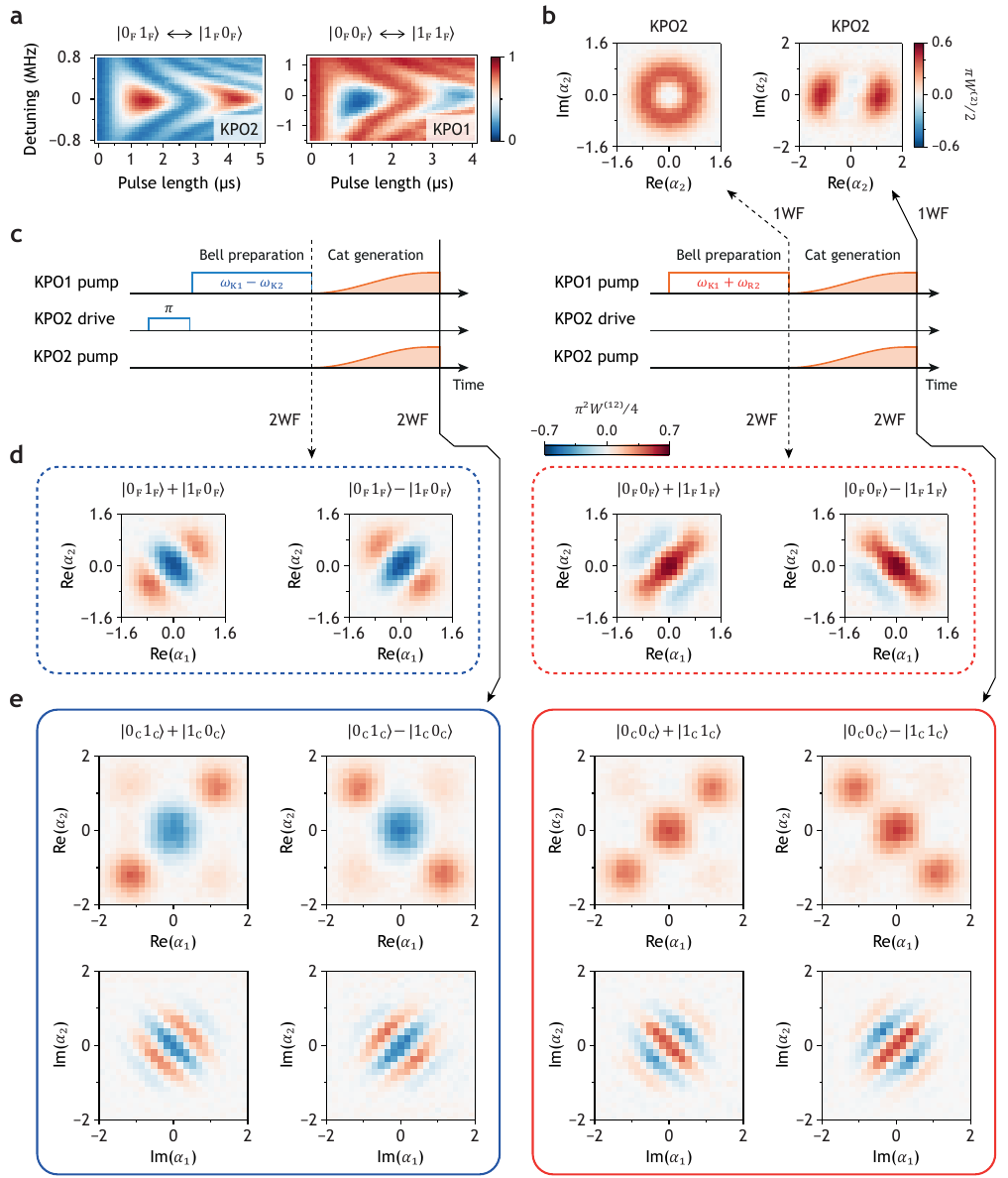}
\caption{\textbf{Converting Bell--Fock states to Bell--Cat states.}
\textbf{a} Rabi oscillations for Bell-preparation pulse.
Regarding Rabi oscillations associated with $\ket{0_\textrm{F}1_\textrm{F}} \leftrightarrow \ket{1_\textrm{F}0_\textrm{F}}$ transitions, the colour represents the population of the $\ket{0_\textrm{F}}$ state of KPO2 and zero detuning corresponds to the frequency $\omega_\textrm{K1}-\omega_\textrm{K2}$.
As for Rabi oscillations associated with $\ket{0_\textrm{F}0_\textrm{F}} \leftrightarrow \ket{1_\textrm{F}1_\textrm{F}}$ transitions, the colour represents the population of the $\ket{0_\textrm{F}}$ state of KPO1 and zero detuning corresponds to the frequency $\omega_\textrm{K1}+\omega_\textrm{K2}-\Delta_\textrm{AC}$, where $\Delta_\textrm{AC}$ is an AC Stark-like frequency shift whose value is 21 MHz in this measurement.
\textbf{b} Measured one-mode Wigner function (1WF) of Bell--Fock and Bell--Cat states.
\textbf{c} Pulse sequences for Bell--Fock state preparation and Bell--Cat state generation.
The amplitude and length of pulses are not to scale.
\textbf{d},\textbf{e}
Measured two-mode Wigner function (2WF) for Bell--Fock (d) and Bell--Cat (e) states.
In Re--Re plots, $\textrm{Im}(\alpha_1)=\textrm{Im}(\alpha_2)=0$,
whereas in Im--Im plots, $\textrm{Re}(\alpha_1)=\textrm{Re}(\alpha_2)=0$.
The colour represents the joint number parity.
}
\label{fig:map}
\end{figure*}

We first prepare all four types of Bell--Fock state,
$\ket{0_\textrm{F} 0_\textrm{F}} \pm \ket{1_\textrm{F} 1_\textrm{F}}$ and 
$\ket{0_\textrm{F} 1_\textrm{F}} \pm \ket{1_\textrm{F} 0_\textrm{F}}$.
Subsequent two-photon pumping to each KPO converts the Bell--Fock state into the same type of Bell--Cat state;
for instance, from $\ket{0_\textrm{F} 0_\textrm{F}} + \ket{1_\textrm{F} 1_\textrm{F}}$
to $\ket{0_\textrm{C} 0_\textrm{C}} + \ket{1_\textrm{C} 1_\textrm{C}}$ (see Fig.~\ref{fig:map}c for the pulse sequence).
This approach relies on the fundamental property of entanglement, namely, that ``entanglement is preserved under local unitary operations'' \cite{plenio}.

The Bell--Fock state is prepared by activating the interaction between the KPOs by applying a parametric pulse with either the frequency $\omega_\textrm{K1}+\omega_\textrm{K2}$ or $\omega_\textrm{K1}-\omega_\textrm{K2}$ to the pump ports \cite{kwon}.
A parametric pulse with each frequency induces the transitions between $\ket{0_\textrm{F}0_\textrm{F}}$ and $\ket{1_\textrm{F}1_\textrm{F}}$, and between $\ket{0_\textrm{F}1_\textrm{F}}$ and $\ket{1_\textrm{F}0_\textrm{F}}$ based on the three-wave mixing capability of our KPOs.
Using such transitions, we can create states $\ket{0_\textrm{F} 0_\textrm{F}} + \textrm{e}^{\textrm{i}\phi_\textrm{s}}\!\ket{1_\textrm{F} 1_\textrm{F}}$ and $\ket{0_\textrm{F} 1_\textrm{F}} + \textrm{e}^{\textrm{i}\phi_\textrm{d}}\!\ket{1_\textrm{F} 0_\textrm{F}}$, where the phases $\phi_\textrm{s}$ and $\phi_\textrm{d}$ are determined by the phase of the parametric pulse (virtual Z gate).
We refer to this pulse as the Bell-preparation pulse.
Rabi oscillations associated with the Bell-preparation pulse are shown in Fig.~\ref{fig:map}a.

For the full characterization of such entangled quantum states, we measured the two-mode Wigner functions (2WFs) (Fig.~\ref{fig:map}d,e) \cite{kim2000,wang2016} because the one-mode Wigner functions (1WFs) cannot provide information on entanglement---all Bell states show the same 1WF, which is identical to that of the fully mixed state (Fig.~\ref{fig:map}b).
The 2WFs of the target Bell--Fock and Bell--Cat states are shown in Supplementary Fig.~3.

We observe all essential features in the 2WF of Bell--Cat states (Fig.~\ref{fig:map}e).
Firstly, in the Re--Re plots with $\Im(\alpha_i)=0$ ($i=1,2$), two red circles aligned diagonally indicate the correlation between the two KPOs, similar to the results in Ref.~\cite{yamaji2023}.
The alignment direction of the red circles represents the sign of the superposition.
The colour of the centre circle, which represents the joint number parity, indicates the type of Bell state;
for instance, $\ket{0 1} + \textrm{e}^{\textrm{i}\phi}\ket{1 0}$ shows a blue centre regardless of whether the basis is Fock or cat.
Secondly, the interference pattern in the Im--Im plot with $\Re(\alpha_i)=0$ demonstrates that the correlation is of quantum nature.

Note that the patterns in Fig.~\ref{fig:map}d,e illustrate how the 2WFs of Bell--Fock states evolve to those of Bell--Cat states:
As the pump amplitude increases, the pattern in Fig.~\ref{fig:map}d elongates along the diagonal axis, eventually resembling the Re--Re plots in Fig.~\ref{fig:map}e.
Regarding the Im--Im plots of Bell--Fock states, those of $\ket{0_\textrm{F} 1_\textrm{F}} \pm \ket{1_\textrm{F} 0_\textrm{F}}$ are identical to the Re--Re plots,
whereas the Im--Im plot of $\ket{0_\textrm{F} 0_\textrm{F}} \pm \ket{1_\textrm{F} 1_\textrm{F}}$ matches the Re--Re plot of $\ket{0_\textrm{F} 0_\textrm{F}} \mp \ket{1_\textrm{F} 1_\textrm{F}}$.
The Im--Im plots in Fig.~\ref{fig:map}e can be interpreted as a compressed version of the plots in Fig.~\ref{fig:map}d along the diagonal axis.
These 2WF patterns show the profound connection between quantum correlations in the Bell--Fock and Bell--Cat states.

The fidelity between the target and measured Bell--Fock states is $0.81\pm 0.01$ (with the error representing the standard deviation);
for the Bell--Cat states, the fidelity is $0.60\pm0.04$.
This value would be 0.25 for completely mixed cat states.
These fidelities were obtained by reconstructing the density matrix from the measured one- and two-mode Wigner functions (see Methods).

The most notable difference between the 2WFs of the measured and target Bell–Cat states is that the measured states show weaker contrast in the centre circle of the Re--Re plots with $\Im(\alpha_i)=0$ ($i=1,2$) and in the interference pattern of the Im--Im plots with $\Re(\alpha_i)=0$ compared to the target states.
(The 2WFs for the target states are shown in Supplementary Fig.~3.)
The primary sources of this difference are thermal excitation and relaxations, such as single-photon loss and dephasing \cite{wang2016,zhao2017}.
Thermal excitation sets an upper limit on fidelity:
if we start from a thermal state rather than a vacuum state, the fidelity is inherently limited, even in the absence of relaxations and with perfect control.
In our case, this upper bound is 0.86, which is the fidelity between the vacuum state and the tensor-product thermal states of both KPOs.

Approximately 0.13 of the fidelity is lost due to relaxations, based on simulations of our Bell–Fock state preparation and cat generation using the Lindblad master equation.
For the cat generation process, dephasing is not considered because low-frequency noise does not affect the fidelity;
the cat states in KPOs are protected by the energy gap \cite{puri2017a, puri2019}.
Experimental evidence suggests that the primary source of relaxation for cat states in a KPO is single-photon loss \cite{grimm2020, catGen, frattini2024}.

Assuming $T_1=T_2=100$ \unit{\micro\second} \cite{place2021, wang2022, kono2024, biznarova2024} and a thermal photon number of 0.01, the fidelity of the Bell-Fock states is approximately 0.96, while the Bell-Cat state fidelity can reach about 0.93, as simulated by the Lindblad master equation.
The main source of remaining infidelity arises from population leakage from the computational subspace due to the small Kerr coefficient.
This issue can be mitigated by using a counterdiabatic or numerically optimized pulse \cite{xue2022, goto2019a, motzoi2009}.
(In this work, a counterdiabatic pulse was not used, unlike in our previous work \cite{catGen}.
See Methods for more information.)

The final fidelity from the simulation, accounting for all these error sources, is approximately 0.68, which is reasonably close to our experimental result.
For more details on the simulation, see Sec.~4A of Supplementary Information.

\subsection{Two-cat gate operation}
\label{sec:gate}

\begin{figure*}
\centering
\includegraphics[scale=0.95]{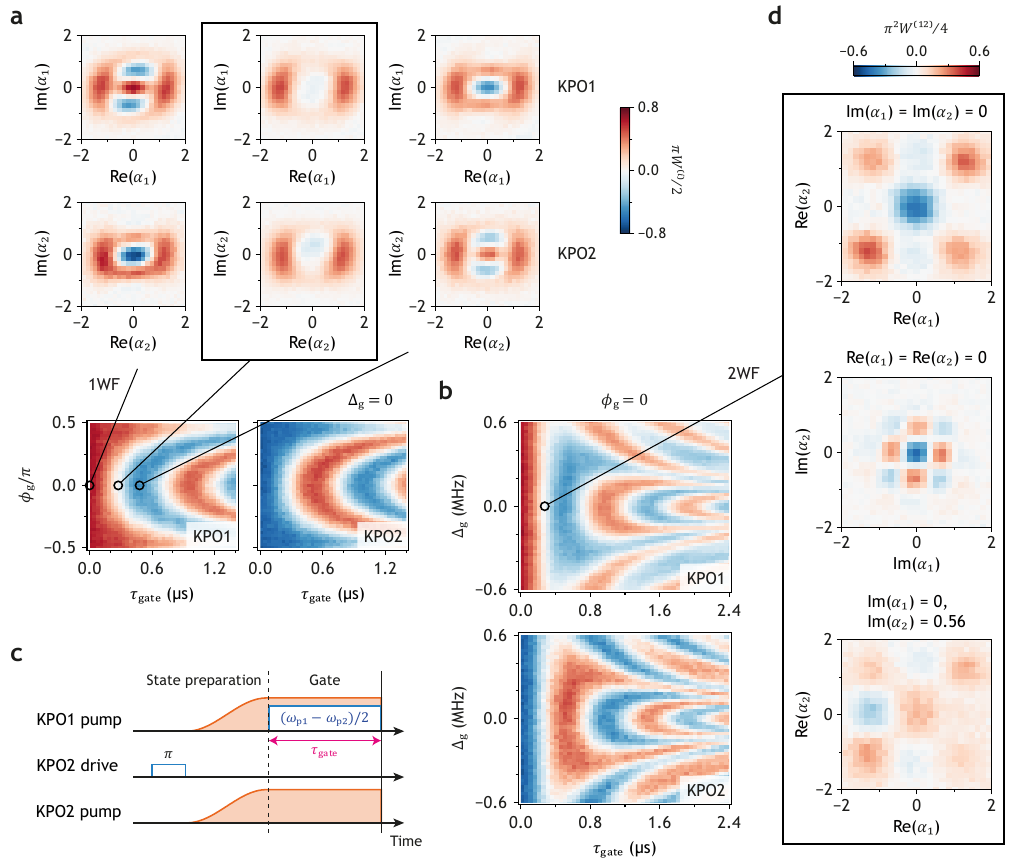}
\caption{\textbf{Two-cat gate operation.}
\textbf{a,b} Two-cat Rabi oscillations between $\ket{0_\textrm{C} 1_\textrm{C}}$ and $\ket{1_\textrm{C} 0_\textrm{C}}$.
The colours represent the number parity of each KPO.
$\phi_\textrm{g}$ and $\Delta_\textrm{g}$ represent the phase and detuning of the gate pulse, respectively.
Zero detuning ($\Delta_\textrm{g}=0$) means that the frequency of the gate pulse is equal to $(\omega_\textrm{p1}-\omega_\textrm{p2})/2$.
One-mode Wigner functions at times corresponding to no gate (0 ns), $\sqrt{\textrm{iSWAP}}$ gate (275 ns), and iSWAP gate (480 ns) are shown above.
\textbf{c} Pulse sequence for the two-cat Rabi.
\textbf{d} Two-mode Wigner functions of the KPO state after the $\sqrt{\textrm{iSWAP}}$ gate.
All Wigner functions showing the results of the $\sqrt{\textrm{iSWAP}}$ gate are enclosed in black frames.
}
\label{fig:gate}
\end{figure*}

One notable feature of this KPO system is that the same type of parametric pulse can be used for two-qubit gate operations in both Fock- and cat-state encodings \cite{chono2022}.
In this work, a parametric pulse---referred to as the gate pulse---with a frequency of $(\omega_\textrm{p1}-\omega_\textrm{p2})/2$ was employed for the gate operation.
If $\Delta_1=\Delta_2$ in Eq.~\eqref{eq:twoKPO}, the gate pulse frequency matches that used to prepare the Bell–Fock state.
Thus, the schemes for creating Bell–Fock and Bell–Cat states are quite similar, with one key difference:
the gate pulse at $(\omega_\textrm{p1}-\omega_\textrm{p2})/2$ induces not only $\ket{0_\textrm{C} 1_\textrm{C}} \leftrightarrow \ket{1_\textrm{C} 0_\textrm{C}}$ transitions but also $\ket{0_\textrm{C} 0_\textrm{C}} \leftrightarrow \ket{1_\textrm{C} 1_\textrm{C}}$ transitions, the latter of which are forbidden in the Fock-state basis but enabled by the presence of the pump.

Below, we explicitly outline the three-wave mixing processes that facilitate this gate operation.
Each term represents the frequency of a photon created or annihilated during the process;
for example, $\omega_\textrm{p2}$ corresponds to the pump for KPO2 and $(\omega_\textrm{p1}-\omega_\textrm{p2})/2$ to the gate pulse.
\begin{align*}
&\textrm{The }\ket{0_\textrm{C} 1_\textrm{C}} \leftrightarrow \ket{1_\textrm{C} 0_\textrm{C}}\textrm{ transitions: } \\ 
&\textrm{(i) } \omega_\textrm{p2} \rightarrow \tfrac{\omega_\textrm{p2}}{2} + \tfrac{\omega_\textrm{p2}}{2}, 
\textrm{ (ii) }
\tfrac{\omega_\textrm{p2}}{2} + \tfrac{\omega_\textrm{p1}-\omega_\textrm{p2}}{2}
\rightarrow
\tfrac{\omega_\textrm{p1}}{2}. \\
&\textrm{The }\ket{0_\textrm{C} 0_\textrm{C}} \leftrightarrow \ket{1_\textrm{C} 1_\textrm{C}}\textrm{ transitions: } \\
&\textrm{(i) }
\omega_\textrm{p2} + \tfrac{\omega_\textrm{p1}-\omega_\textrm{p2}}{2}
\rightarrow
\tfrac{\omega_\textrm{p1}+\omega_\textrm{p2}}{2}, 
\textrm{ (ii) }
\tfrac{\omega_\textrm{p1}+\omega_\textrm{p2}}{2}
\rightarrow
\tfrac{\omega_\textrm{p1}}{2} + \tfrac{\omega_\textrm{p2}}{2}.
\end{align*}
Here, the processes (i) and (ii) occur sequentially.

Based on the three-wave mixing processes described above, particularly process (ii), the working principle of our two-cat gate operation can be understood as follows.
We begin with the X gate, which can be implemented using a single-photon drive with the frequency $\omega_\textrm{p1}/2$ or $\omega_\textrm{p2}/2$ \cite{catGen}, as the X gate changes the parity of the state and the cat states are already shaped by the interplay between the Kerr nonlinearity and the two-photon pump.
In our two-cat gate, the gate pulse enables the two KPOs to exchange a single photon (for the $\ket{0_\textrm{C} 1_\textrm{C}} \leftrightarrow \ket{1_\textrm{C} 0_\textrm{C}}$ transitions) or absorb/emit a single photon simultaneously (for the $\ket{0_\textrm{C} 0_\textrm{C}} \leftrightarrow \ket{1_\textrm{C} 1_\textrm{C}}$ transitions).
Such processes result in two correlated X gates acting on each KPO, implementing the following gate operation $\hat{U}_\textrm{G}$:
\begin{equation}\label{eq:gateMatrix}
\hat{U}_\textrm{G} = \frac{1}{\sqrt{2}}
\begin{pmatrix}
1 				&	0				&	0				&	\textrm{i} \\
0			 	&	1	 			&	\textrm{i}	&	0 \\
0 				&	\textrm{i}	&	1				&	0 \\
\textrm{i}	&	0				&	0				&	1 
\end{pmatrix}.
\end{equation}
Within the $\ket{0_\textrm{C} 1_\textrm{C}}$ and $\ket{1_\textrm{C} 0_\textrm{C}}$ subspace, the gate operation remains identical to the $\sqrt{\textrm{iSWAP}}$ gate.
For simplicity, we refer to it as the $\sqrt{\textrm{iSWAP}}$ gate in this work.
(This work does not explore $\ket{0_\textrm{C} 0_\textrm{C}} \leftrightarrow \ket{1_\textrm{C} 1_\textrm{C}}$ transitions.
A complete characterization of the gate operation, including this aspect, is left for future study.)

Note that Eq.~\eqref{eq:gateMatrix} corresponds to the $\hat{R}_\textrm{ZZ}$ gate in the coherent-state basis (see Sec.~6 of the Supplementary Information for the derivation).
This has two key implications:
Since the $\hat{R}_\textrm{ZZ}$ gate is part of the universal gate set, Eq.~\eqref{eq:gateMatrix} is as well.
Additionally, the $\hat{R}_\textrm{ZZ}$ gate---and therefore Eq.~\eqref{eq:gateMatrix}---preserves the biased-noise property of KPOs, as discussed in Ref.~\cite{puri2020}.

We observe the Rabi-like oscillations in the parity of each KPO, which we call the two-cat Rabi, as a function of the phase and the detuning of the parametric pulse (Fig.~\ref{fig:gate}a,b).
Here, the gate phase $\phi_\textrm{g}$ is the phase relative to the pumps, and the gate detuning $\Delta_\textrm{g}$ is the detuning from $(\omega_\textrm{p1}-\omega_\textrm{p2})/2$.
For this measurement, we first prepare $\ket{0_\textrm{F}1_\textrm{F}}$ and convert it to $\ket{0_\textrm{C}1_\textrm{C}}$ by applying the pumps.
Then, we apply the gate pulse, in addition to the pumps, as shown in Fig.~\ref{fig:gate}c.

Note that the two KPOs exhibit the same two-cat Rabi oscillations but with opposite parities.
From the simulation, we determined the gate amplitude to be 2.96 MHz (see Supplementary Fig.~5a and its caption for details).
One-mode Wigner functions show that during the Rabi oscillations, the state evolves from 
$\ket{0_\textrm{C}1_\textrm{C}}$ (no gate) to $\ket{1_\textrm{C}0_\textrm{C}}$ (iSWAP).
To determine the intermediate quantum state between these two points, a 2WF measurement with an additional offset in displacement is required.
This is because 1WF and the Re--Re (Im--Im) plot in 2WF without an additional displacement along the imaginary (real) axes cannot distinguish the following three states:
$\ket{0_\textrm{C}1_\textrm{C}} \pm \textrm{i}\ket{1_\textrm{C}0_\textrm{C}}$, which are the states after the $\sqrt{\textrm{iSWAP}}$ gate, and the mixture of $\ket{0_\textrm{C}1_\textrm{C}}$ and $\ket{1_\textrm{C}0_\textrm{C}}$. 
The Re--Re plot with an additional offset shows that the state is $\ket{0_\textrm{C}1_\textrm{C}} - \textrm{i}\ket{1_\textrm{C}0_\textrm{C}}$ (the plot at the bottom of Fig.~\ref{fig:gate}d), confirming that the two-cat gate operation is the $\sqrt{\textrm{iSWAP}}$ gate (see Supplementary Fig.~3).

The $\sqrt{\textrm{iSWAP}}$ gate time, 275 ns, is significantly faster than recent implementations of similar SWAP gate operations on bosonic modes \cite{gao2019, chapman2023}.
This short gate time is possible because the beam-splitter interaction is inherently built into the Hamiltonian [Eq.~\eqref{eq:twoKPO}], and the KPO system enables us to adopt schemes for gate operations in Fock-state encoding.
The primary limitations on our gate time are the AC Stark-like frequency shift induced by the gate pulse above a certain amplitude threshold, which would introduce unwanted Z-gate operations, and the small cat size.
Additionally, Ref.~\cite{chono2022} suggested implementing Eq.~\eqref{eq:gateMatrix} using the frequency $(\omega_\textrm{p1}+\omega_\textrm{p2})/2$, as we demonstrated in Fig.~\ref{fig:map} for the preparation of Bell--Fock states.
We did not pursue this approach because the amplitude threshold for the AC Stark-like frequency shift is almost zero at $(\omega_\textrm{p1}+\omega_\textrm{p2})/2$.
Therefore, suppressing the AC Stark-like frequency shift at the circuit design level and increasing the cat size will enable faster gate operations and enhance functionalities.

Similarly to the Bell--Fock state preparation, the sign of the superposition can be flipped by adding $\pi$ in the phase of the two-cat gate pulse.
Unlike the conversion from the Bell--Fock to Bell--Cat states, however, we cannot create a Bell–Cat state with an arbitrary phase.
The reason is that once the pumps are turned on, the pump phase becomes the reference phase.
Consequently, we can no longer use the virtual Z gate as implied in Fig.~\ref{fig:gate}a.
In other words, $\phi_\textrm{d}$ in $\ket{0_\textrm{F} 1_\textrm{F}} + \textrm{e}^{\textrm{i}\phi_\textrm{d}}\!\ket{1_\textrm{F} 0_\textrm{F}}$ can only take values of $(n+1/2)\pi$, where $n$ is an integer.
Thus, the Bell--Cat state we create in this work using the two-cat gate is limited to $\ket{0_\textrm{C}1_\textrm{C}} \pm \textrm{i}\ket{1_\textrm{C}0_\textrm{C}}$, which is exactly what is expected from the $\sqrt{\textrm{iSWAP}}$ gate.

The gate-detuning dependence of the two-cat Rabi exhibits the characteristic pattern observed in cat Rabi oscillations for the X gate \cite{catGen}.
This suggests that, as pointed out in Ref.~\cite{catGen}, when mapping the dynamics of cat states to that of interacting two-level qubits, two tones with opposite gate detuning are required.
In such a two-level qubit system, the same pattern can be reproduced by modulating the coupling constant with two frequencies, $\omega_\textrm{g}$ and $2(\omega_\textrm{q1}-\omega_\textrm{q2})-\omega_\textrm{g}$, where $\omega_{\textrm{q}i}$ is the transition frequency of the two-level qubit$i$ ($i=1,2$).
In this case, zero detuning corresponds to $\omega_\textrm{g} = \omega_\textrm{q1}-\omega_\textrm{q2}$.
For further discussion and simulation results, see Sec.~5 of Supplementary Information.

The fidelity of the $\ket{0_\textrm{C} 1_\textrm{C}} \pm \textrm{i}\ket{1_\textrm{C} 0_\textrm{C}}$ states is $0.61\pm0.03$, which is almost identical to that achieved by the conversion from Bell--Fock to Bell--Cat states.
This result is not surprising because, although the pulse length of the $\sqrt{\textrm{iSWAP}}$ gate on the cat states (275 ns) is less than half that of the Bell-preparation pulse (730 ns),
the contrast of the two-cat Rabi oscillations attenuates faster than that of the Rabi oscillations used for the Bell--Fock state preparation (compare Fig.~\ref{fig:gate}b and the left plot of Fig.~\ref{fig:map}a). 
More quantitatively, the decay time of the two-cat Rabi oscillation is 3 \unit{\micro\second} for both KPOs, whereas that of the Rabi oscillations used in Bell--Fock state preparation is longer than 10 \unit{\micro\second}.
The simulation using the Lindblad master equation suggests that the photon lifetime of both KPOs to reproduce the data in Fig.~\ref{fig:gate}b is 10 \unit{\micro\second} (see Supplementary Fig.~5a).
This photon lifetime falls within the observed range (Supplementary Table~1).
Thus, the main sources of infidelity in this case are also thermal excitation and relaxations.

The fidelity of our Bell-Cat states is somewhat lower than that reported in previous works, which was 0.74 \cite{gao2019, chapman2023}.
This difference is likely due to the low thermal population ($\lesssim\verythinspace$0.01) in the earlier studies.
In fact, the 0.13 difference in fidelity between our work and previous studies closely matches the infidelity attributed to thermal excitation, as discussed in Sec.~\ref{sec:conversion}.
We expect the performance of our gate operation to be comparable to that of previous demonstrations.

Overall, our gate operation is faster and significantly simpler, requiring only a single square pulse to implement the $\sqrt{\textrm{iSWAP}}$ gate, while achieving comparable performance.
Furthermore, there is considerable potential for improving fidelity by suppressing the thermal population, enhancing the relaxation times of KPOs, or optimizing the gate pulse shape.

Lastly, we point out that the same $\sqrt{\textrm{iSWAP}}$ gate operation can be performed between cat states with different mean photon numbers.
This property may provide significant flexibility when constructing a KPO-based quantum network, particularly for the scheme developed in Refs.~\cite{zhong2021, qiu2023}.
The simulation results can be found in Supplementary Fig.~5c.

\section{Discussion}

To summarize, we demonstrate two intuitive methods for entangling cat states by adopting a DV--CV hybrid approach.
This hybridization is achieved through Hamiltonian engineering, combining moderate Kerr nonlinearity and two-photon pumping.
It enables coherent treatment of Bell--Fock and Bell--Cat states, facilitating gate operations directly on the cat basis without the need for ancilla qubits or individual Fock state control.
One consequence is the entanglement-preserving conversion from Bell--Fock to Bell--Cat states.
The other is the fast and simple $\sqrt{\textrm{iSWAP}}$ gate operation on the cat states, thereby completing the demonstration of a universal quantum gate set.
Therefore, our superconducting planar KPO system is not only a promising quantum information processing unit but also a potent platform for DV--CV hybridization.

We suggest several future research directions extending this work.
First, we can construct quantum networks in the cat basis.
Note that our methods are compatible with previously demonstrated quantum network constructions in the Fock basis \cite{kurpiers2018, zhong2021, qiu2023}.
This means we can create more complex entangled states, such as Greenberger–Horne–Zeilinger or cluster states, in the cat basis simply by replacing transmon/Xmon qubits with KPOs and converting the basis from Fock to cat states.
This approach will significantly reduce the complexity of constructing quantum networks using bosonic modes.
We can also create travelling entangled-cat states by coupling our system to transmission lines \cite{goto2019a}.
Combining Hamiltonian engineering with dissipation engineering may enable us to create highly coherent cat states \cite{gautier2022,gravina2023,marquet2024,reglade2024}.
Finally, employing other multiphoton pumps may open new possibilities \cite{svensson2017, svensson2018, chang2020, zhang2017a, zhang2019, tadokoro2020, gosner2020, lang2021, arndt2022, miganti2023, iachello2023, guo2024, mora2024}, such as exploring condensed matter physics in time crystals \cite{guo2013, guo2020, sacha} and autonomous quantum error correction \cite{kwon2022}.

\section{Methods}

\subsection{Cat state generation}

As mentioned in the main text, the ramping time of the pump for cat-state generation is 1 \unit{\micro\second}.
This ramping time is much longer than that in our previous work (300 ns) \cite{catGen} because the counterdiabatic pulse did not work.
We believe that the reason is the reduction in Kerr coefficient from about 3 MHz to 2 MHz after ramping up
the pump (see Supplementary Table 1), whereas the Kerr coefficient in Ref.~\cite{catGen} increases slightly from 2.86 MHz to 3.13 MHz.

During the ramping, we change the pump frequency, i.e., chirp the pump pulse, for two reasons:
One is to compensate for unwanted AC Stark-like frequency shifts in $2\omega_\textrm{K1}$ and $2\omega_\textrm{K2}$, which are approximately $-$10 MHz at the target pump amplitude \cite{catGen}.
The other reason is that the pump detuning must start from zero and then approach the target value adiabatically to create high-fidelity cat states.

\subsection{Wigner-function measurements}

\begin{figure}
\centering
\includegraphics[scale=0.95]{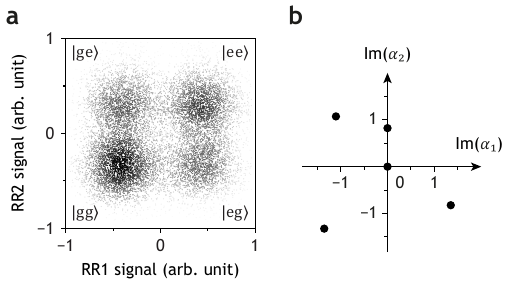}
\caption{\textbf{Configurations for measuring two-mode Wigner functions.}
\textbf{a} Typical single-shot results.
RR stands for ``readout resonator''.
\textbf{b} Coordinates of offset displacement for Re--Re plots of a two-mode Wigner function.
}
\label{fig:fitting}
\end{figure}

The one-mode Wigner function of the KPO is given by \cite{royer1977}
\begin{equation} \label{eq:1wigner}
W^{(i)}(\alpha_i) = 
\frac{2}{\pi} \Tr[\hat{D}^\dagger(\alpha_i) \rho^{(i)} \hat{D}(\alpha_i) \hat{\Pi}^{(i)}],
\end{equation}
where
$\hat{D}(\alpha_i) = \textrm{exp}\big(\alpha_i \hat{a}_i^\dagger - \alpha_i^* \hat{a}_i \big)$ is the displacement operator,
$\hat{\Pi}^{(i)} = \textrm{exp}\big(\textrm{i}\pi \hat{a}_i^\dagger\hat{a}_i \big)$ is the photon-number parity operator, and
$\rho^{(i)}$ is the density matrix of KPO$i$ $(i=1,2)$.
Similarly, the two-mode Wigner function is given by \cite{wang2016}
\begin{align}
&W^{(12)}(\alpha_1,\alpha_2) \nonumber\\
&\, = \frac{4}{\pi^2} 
\Tr[\hat{D}^\dagger(\alpha_2)\hat{D}^\dagger(\alpha_1) \rho^{(12)} 
\hat{D}(\alpha_1) \hat{D}(\alpha_2) \hat{\Pi}^{(12)}]  \\
&\, = \frac{4}{\pi^2} \langle \hat{\Pi}^{(12)}(\alpha_1,\alpha_2) \rangle, \label{eq:2wigner}
\end{align}
where
$\rho^{(12)}$ is the density matrix of the two-KPO system and
$\hat{\Pi}^{(12)} = \hat{\Pi}^{(1)}\hat{\Pi}^{(2)}$ is the joint parity of the two KPOs.
This operator can be measured by the joint probabilities of the transmons being in their ground/excited state $P_{jk}$ ($j,k \in \{\textrm{g},\textrm{e}\}$) \cite{kim2000}:
\begin{equation}
\langle \hat{\Pi}^{(12)}(\alpha_1,\alpha_2) \rangle = 
P_\textrm{ee} + P_\textrm{gg} - P_\textrm{eg} - P_\textrm{ge}.
\end{equation}
In the experiment, this was accomplished by fitting the single-shot readout data (Fig.~\ref{fig:fitting}a) with a two-dimensional Gaussian function for all pixels of the Wigner function plots.

The two-mode Wigner functions of the target and simulated states in Supplementary Information are obtained using the Cahill--Glauber formula \cite{cahill1969a,cahill1969b,miranowicz2023}:
\begin{align}
W^{(12)} &(\alpha_1,\alpha_2) \nonumber\\
& = \frac{4}{\pi^2}
\Tr[\rho^{(12)} \hat{T}(\alpha_1)\hat{T}(\alpha_2)] \\
\begin{split}
& = \frac{4}{\pi^2}
\sum_{\{n_i\}=0}^{N_i} \sum_{\{m_i\}=0}^{N_i} \prod_{i=1}^{2}
\langle n_i | \hat{T} (\alpha_i) | m_i \rangle \\
&\quad\, \times \langle \{m_i\} | \rho^{(12)} | \{n_i\} \rangle,
\end{split}
\end{align}
where
$\hat{T}$ is the complex Fourier transform of the displacement operator, and
$N_i$ is the dimension of the Hilbert space of KPO$i$.
For $m_i \geq n_i$,
\begin{equation}
\begin{split}
\langle n_i | \hat{T} (\alpha_i) | m_i \rangle =\,& 
\sqrt{\frac{n_i !}{m_i !}} (-1)^{n_i} (2\alpha_i^*)^{\delta_i} \\
& \times L_{n_i}^{(\delta_i)}(4|\alpha_i|^2) \exp(-2|\alpha_i|^2),
\end{split}
\end{equation}
where $\delta_i \equiv m_i-n_i$ and $L_{n_i}^{(\delta_i)}(x)$ are the associated Laguerre polynomials.
For $m_i < n_i$, we can use the following property:
\begin{equation}
\langle n_i | \hat{T} (\alpha_i) | m_i \rangle = \langle m_i | \hat{T} (\alpha_i^*) | n_i \rangle.
\end{equation}

\subsection{Density-matrix reconstruction}

A two-mode Wigner function is a four-dimensional function.
Since our signal-to-noise ratio is marginal, as shown in Fig.~\ref{fig:fitting}a, collecting such a large data set---$13 \times 13 \times 13 \times 13$ pixels, for example---is impractical.
Instead, we measured 10 two-dimensional plots, each of which has $17\times 17$ pixels in the range $-1.6 \leq \alpha_i \leq 1.6$ ($i=1,2$).
Among these 10 plots, half are Re--Re plots with imaginary offset displacements and the other half are Im--Im plots with real offset displacements.
For Re--Re plots, the imaginary offset displacements are given as follows (Fig.~\ref{fig:fitting}b):
$\{(\Im(\alpha_1),\Im(\alpha_2))\} = \{(0,0)$, $(0,+0.82)$, $(-1.10,+1.07)$, $(-1.35,-1.32)$, $(+1.35,-0.82)\}$.
The same values are used for the real offset displacements for Im--Im plots.
In addition to the 10 plots from the two-mode Wigner function, one-mode Wigner functions of both KPOs were measured, resulting in a total of 12 plots used for density matrix reconstruction.

We found that, with the data set simulated from the target Bell--Cat states, the reconstruction fidelity is $>\verythinspace$0.99.
We also checked the reconstruction fidelity of non-ideal data sets.
For example, we prepared low-quality Bell--Cat states by simulating the Lindblad mater equation with $T_1=10$ \unit{\micro\second} for both KPOs after 2 \unit{\micro\second} waiting; the resulting fidelity between this state and the initial state was 0.57, which is similar to our results.
The reconstruction fidelity from 2WFs of these low-quality states is still $\geq\verythinspace$0.98.

For Bell--Fock states, the dimensions of the Hilbert space were set to $3 \times 3$.
For Bell--Cat states, the dimensions of the Hilbert space were set to $8 \times 8$ because, for the ideal Bell--Cat states with $P/K = 1$ and 1 MHz of pump detuning, the occupation probability at $\ket{\geq\!8}$ is less than $10^{-4}$.

The algorithm for reconstruction followed the idea from Refs.~\cite{ahmed2021a, ahmed2021b}, which use gradient descent to reconstruct a density matrix with a projection step.
A loss function between the measured data and that obtained from an estimated density matrix is minimized to obtain the reconstructed density matrix starting from a random initialization.
We simplified the method to directly apply gradient descent (Adam \cite{adam}) on a matrix $T$, that is projected to construct an estimate of the physical density matrix using the Cholesky decomposition.
At each gradient-descent step, the loss function is minimized followed by a projection step where the matrix $T$ is converted to a lower triangular matrix with real-valued diagonal elements by discarding the upper-triangular part and making the diagonal real.
This step allows us to obtain a density matrix $\rho = \frac{T^{\dagger} T}{\text{Tr}(T^{\dagger} T)}$ that is guaranteed to be physical.
The Python libraries used were QuTiP \cite{qutip1, qutip2}, NumPy \cite{numpy}, and JAX \cite{jax}.

\bigskip

\textbf{Data availability:} All data are available in the main text or in the supplementary information.

\bigskip

\textbf{Acknowledgments:}
The authors thank Adam Miranowicz, Tsuyoshi Yamamoto, Shiro Saito, Atsushi Noguchi, Shotaro Shirai, and Yoshiki Sunada for their interest in this project and helpful discussion.
We also thank
Kazumasa Makise of the National Astronomical Observatory of Japan for providing niobium films and
the MIT Lincoln Laboratory for providing a Josephson travelling-wave parametric amplifier.
This work was primarily supported by the Japan Science and Technology Agency (Moonshot R\&D, JPMJMS2067; CREST, JPMJCR1676) and the New Energy and Industrial Technology Development Organization (NEDO, JPNP16007).
These funds were secured by JST.
SA and AFK acknowledge support from the Knut and Alice Wallenberg Foundation through the Wallenberg Centre for Quantum Technology (WACQT).
AFK is also supported by the Swedish Research Council (grant number 2019-03696), the Swedish Foundation for Strategic Research (grant numbers FFL21-0279 and FUS21-0063), and the Horizon Europe programme HORIZON-CL4-2022-QUANTUM-01-SGA via the project 101113946 OpenSuperQPlus100.

\medskip

\textbf{Author contributions:}
SK and JST conceived the project.
SK, DH, TN, HM, and JST designed the details of the experiment.
DH, TN, and SK performed the measurements and data analysis.
TN and SK performed the simulations with contributions from SF.
SW provided theoretical support.
DH and DI wrote the software for the measurements.
SA wrote the code for the density-matrix reconstruction with contributions from AFK and SK.
HM managed the hardware.
SK and TK designed the chip.
TK fabricated the chip.
SK wrote the original draft with contributions from DH and TN.
All authors contributed to the review and editing of the paper.
SK, FY, and JST supervised the project.

\medskip

\textbf{Competing interests:} The authors declare that they have no competing interests.

\balancecolsandclearpage
\includepdf[pages=1]{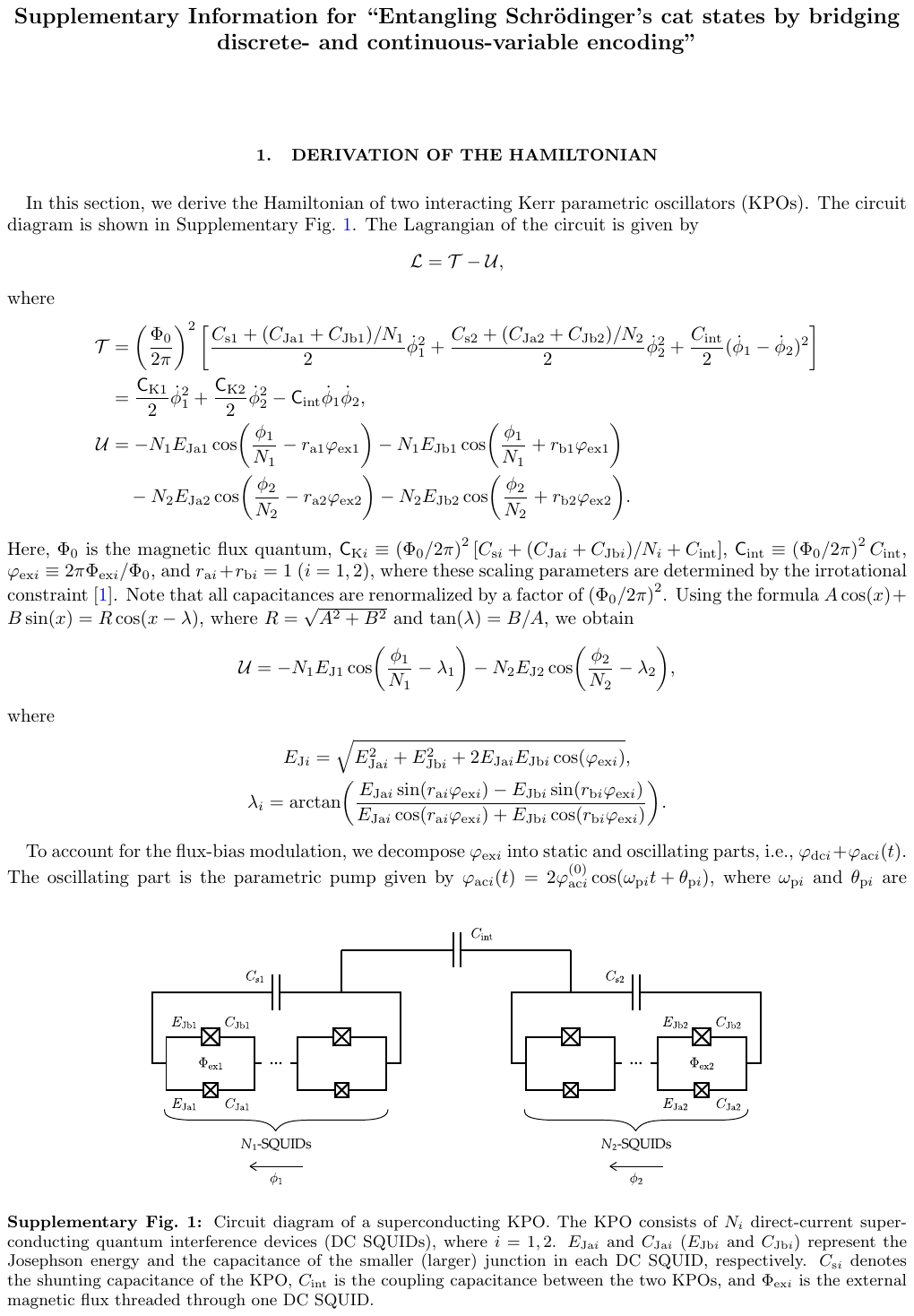}
$ $

\newpage
\includepdf[pages=2]{bellCat_supp.pdf}
$ $
\newpage
\includepdf[pages=3]{bellCat_supp.pdf}
$ $
\newpage
\includepdf[pages=4]{bellCat_supp.pdf}
$ $
\newpage
\includepdf[pages=5]{bellCat_supp.pdf}
$ $
\newpage
\includepdf[pages=6]{bellCat_supp.pdf}
$ $
\newpage
\includepdf[pages=7]{bellCat_supp.pdf}
$ $
\newpage
\includepdf[pages=8]{bellCat_supp.pdf}
$ $
\newpage
\includepdf[pages=9]{bellCat_supp.pdf}
$ $
\newpage
\includepdf[pages=10]{bellCat_supp.pdf}
$ $
\newpage
\includepdf[pages=11]{bellCat_supp.pdf}
$ $
\newpage
\includepdf[pages=12]{bellCat_supp.pdf}
$ $

\end{document}